\documentclass[fleqn,usenatbib]{mnras}

\usepackage{newtxtext,newtxmath}
\usepackage[T1]{fontenc}
\usepackage{ae,aecompl}

\usepackage{graphicx}	% Including figure files
\usepackage{amsmath}	% Advanced maths commands
\usepackage{amssymb}	% Extra maths symbols

\def\msun{\hbox{M$_\odot$}}

\usepackage[dvipsnames]{xcolor}
\usepackage{ulem}
\title[NGC~1978 with MUSE]{Leveraging HST with MUSE: I. Sodium abundance variations within the 2 Gyr-old cluster NGC 1978}

\author[Saracino et al.]{S. Saracino$^{1}$\thanks{E-mail: s.saracino@ljmu.ac.uk}%
, S. Kamann$^{1}$,
C. Usher$^{2}$,
N. Bastian$^{1}$,
S. Martocchia$^{1,3}$,
C. Lardo$^{4}$,
\newauthor
M. Latour$^{5}$,
I. Cabrera-Ziri$^{6}$\thanks{Hubble Fellow},
S. Dreizler$^{5}$,
B. Giesers$^{5}$,
T.-O. Husser$^{5}$,
N. Kacharov$^{7}$
\newauthor
M. Salaris$^{1}$
\\
$^{1}$Astrophysics Research Institute, Liverpool John Moores University, 146 Brownlow Hill, Liverpool L3 5RF, UK\\
$^{2}$Department of Astronomy, Oskar Klein Centre, Stockholm University, AlbaNova University Centre, SE-106 91 Stockholm, Sweden \\
$^{3}$European Southern Observatory, Karl-Schwarzschild-Stra\ss e 2, D-85748 Garching bei M\"unchen, Germany\\
$^{4}$Geneva Observatory, University of Geneva, Maillettes 51, 1290, Sauverny, Switzerland\\
$^{5}$ Institute for Astrophysics, Georg-August-University G\"ottingen, Friedrich-Hund-Platz 1, D-37077 G\"ottingen, Germany\\
$^{6}$Harvard-Smithsonian Center for Astrophysics, 60 Garden Street, Cambridge, MA 02138, USA\\
$^{7}$Max-Planck-Institut f\"ur Astronomie, K\"onigstuhl 17, D-69117 Heidelberg, Germany\\
}
\date{Accepted September 2, 2019. Received August 3, 2019; in original form May 22, 2019}

\pubyear{2020}

\begin{document}
\label{firstpage}
\pagerange{\pageref{firstpage}--\pageref{lastpage}}
\maketitle

\begin{abstract}
Nearly all of the well studied ancient globular clusters (GCs), in the Milky Way and in nearby galaxies, show star-to-star variations in specific elements (e.g., He, C, N, O, Na, Al), known as ``multiple populations" (MPs). However, MPs are not restricted to ancient clusters, with massive clusters down to $\sim2$~Gyr showing signs of chemical variations. This suggests that young and old clusters share the same formation mechanism but most of the work to date on younger clusters has focused on N-variations. Initial studies even suggested that younger clusters may not host spreads in other elements beyond N (e.g., Na), calling into question whether these abundance variations share the same origin as in the older GCs. In this work, we combine HST photometry with VLT/MUSE spectroscopy of a large sample of RGB stars (338) in the Large Magellanic Cloud cluster NGC~1978, the youngest globular to date with reported MPs in the form of N-spreads. By combining the spectra of individual RGB stars into N-normal and N-enhanced samples, based on the ``chromosome map'' derived from HST, we search for mean abundance variations. Based on the NaD line, we find a Na-difference of $\Delta$[Na/Fe]$=0.07\pm0.01$ between the populations. While this difference is smaller than typically found in ancient GCs (which may suggest a correlation with age), this result further confirms that the MP phenomenon is the same, regardless of cluster age and host galaxy. As such, these young clusters offer some of the strictest tests for theories on the origin of MPs.
\end{abstract}


%%%%%%%%%%%%%%%%%%%%%%%%%%%%%%%%%%%%%%%%%%%%%%%%%%

%%%%%%%%%%%%%%%%% BODY OF PAPER %%%%%%%%%%%%%%%%%%
\section{Introduction}
Globular clusters (GCs) host sub-populations of stars with distinctive light element abundance patterns, known as multiple populations (MPs). In particular, some GCs stars show N, Na and Al enhancement along with C and O depletion while others display the abundance ratios seen in field stars of the same metallicity (see \citealt{bastian2018} for a recent review). The origin of these variations is still under debate, as we currently lack a formation mechanism that is able to explain all the observational findings.

A lot of effort has been put so far in the context of MPs from a photometric point of view. The ultraviolet (UV) legacy survey of Galactic GCs (\citealt{piotto2015}, \citealt{nardiello2018}) provided the community with a global census of MP properties for a large sample of the clusters in the Milky Way, using an appropriate HST filter combination (from the UV to the optical) able to distinguish populations of stars with different N and He abundances \citep{milone2017}.

In the last years the same has been done for clusters in the Large and Small Magellanic Clouds (MCs), with the advantage of covering a quite large age range (\citealt{niederhofer2017b,niederhofer2017a,martocchia2018a,martocchia2018b}). This survey established an age limit of $\sim2$~Gyr for the appearance of detectable MPs in massive ($\gtrsim10^{5}$~\msun) star clusters \citep{martocchia2017}. Using a consistent approach, \citet{saracino2019} have demonstrated that, at least from a photometric point of view, the MP phenomenon shows the same properties irrespective of clusters age and/or the environment where they live. This is an interesting result by itself, but it needs to be corroborated spectroscopically, for example for elements like Na and Al, in order to conclusively link the Magellanic Cloud clusters to their ancient Milky Way counterparts.

Recently, \citet{salaris2020} showed that the width of the red giant branch (RGB) related to MPs in clusters is affected by the first dredge up, which causes significant mixing in N, an element that dominates photometric detections of MPs (including the ``chromosome maps'' - \citealt{milone2017}). In particular, the strength of the first dredge up is strongly dependent on the mass (hence age) of the star as it ascends the sub-giant branch (SGB) and RGB. The result of this effect is to make the N-normal stars appear closer to the N-enhanced stars, leading to an underestimation of the N-spread within the cluster at younger ages.

In order to demonstrate the equivalence of MPs in young-intermediate age clusters and the ancient GCs, we now explore elements that are not (or significantly less) affected by the first dredge up, specifically we need to search for Na spreads, a key element in previous spectroscopic studies of the MP phenomenon \citep[e.g.][]{carretta2009, latour2019}.
To do so, we have started a spectroscopic campaign with MUSE (GTO time, PI: S. Kamann), targeting a sample of young and intermediate age clusters in the MCs, with the aim of looking for internal Na, Mg and Al abundance differences.

Here we present the results of our pilot study, focusing on NGC~1978, a massive ($\sim3\cdot 10^5$ $M_{\odot}$, \citealt{westerlund1997}), $\sim$2 Gyr-old cluster \citep{mucciarelli2007} in the Large Magellanic Cloud.  To date, it is the youngest cluster to show evidence of MPs \citep{martocchia2019}.
This paper is structured as follows: in Section \ref{sec:obs} the observational data-sets and reduction techniques are presented. In Section \ref{sec:res} we discuss our results, describing the innovative approach used to treat the spectroscopic data: from the chromosome map as a diagnostic for MPs, to the Na abundance analysis of the N-normal and N-enriched populations within the cluster. In Section \ref{sec:concl} we summarize the most relevant findings and draw our conclusions.
\vspace{-0.6cm}
\section{Observations and data analysis}\label{sec:obs}
\subsection{HST photometry}\label{sec:obs:hst}
\begin{figure}
    \centering
	\includegraphics[width=0.47\textwidth]{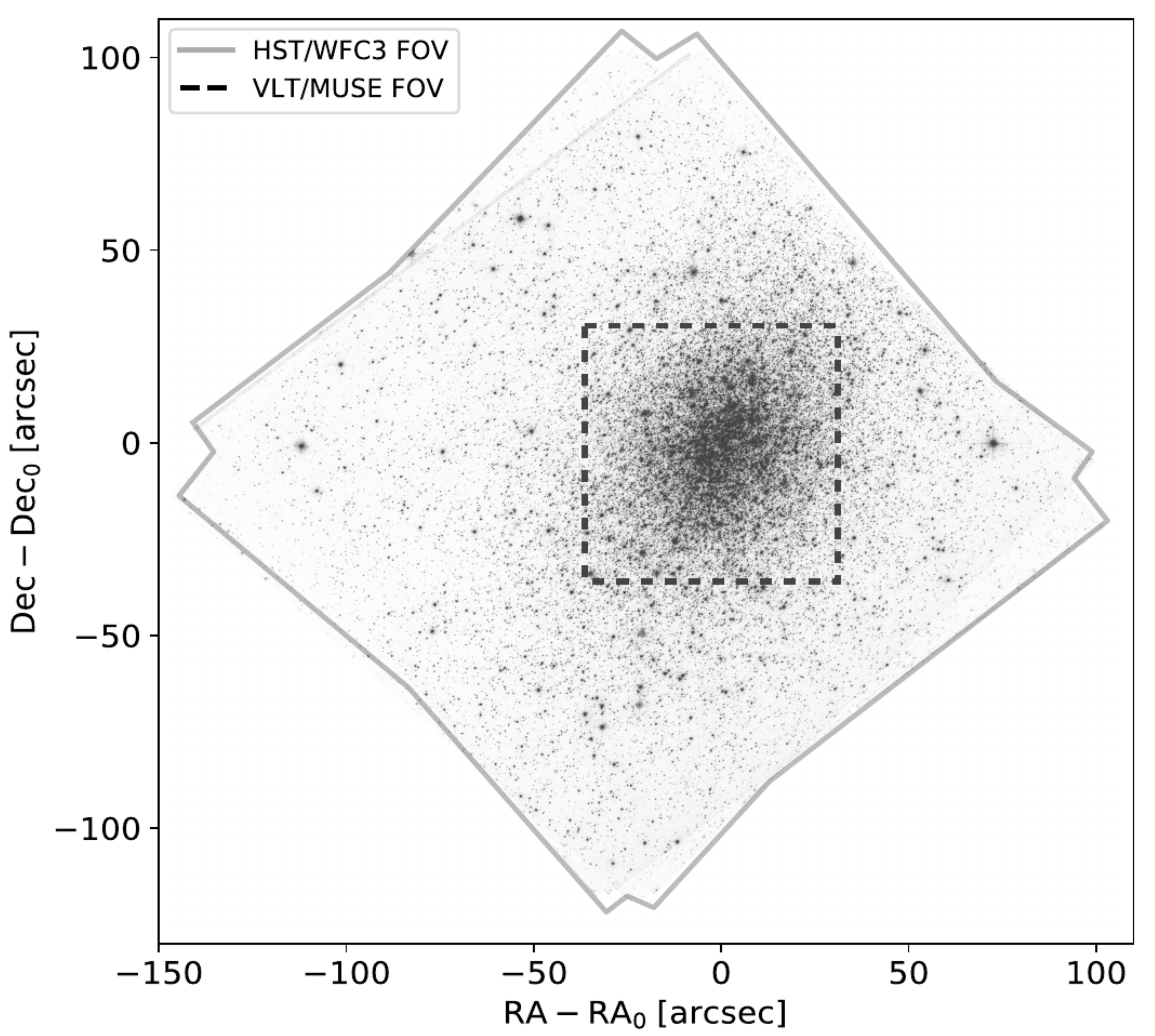}
    \caption{Black and white image of NGC 1978, from the Hubble Space Telescope. The grey solid line refers to the WFC3 field of view covered by the two HST programs used in this study. The black dashed box shows the VLT/MUSE field of view, targeting the innermost part of the cluster. A combined photometric and spectroscopic analysis of NGC~1978 has been performed for the stars within the overlapping fields of view.}
    \label{fig:map}
\end{figure}

This work is based on two HST proposals (GO-14069, GO-15630, PI: N. Bastian) composed of near-UV and optical images obtained through the Wide Field Camera 3 (WFC3) UVIS channel as part of our LMC/SMC survey of star clusters. In particular, observations in the wide filters F275W, F336W, F438W and F814W and in the narrow filter F343N are available for NGC 1978. The field of view covered by these HST observations is shown in Fig.~\ref{fig:map} as a grey solid line, superimposed on a black and white HST image of the cluster.
The photometric analysis of the data has been performed using DAOPHOT IV \citep{stetson1987}, and the cross-correlation software CataXcorr \citep{montegriffo1995}. The procedure adopted is standard and has been extensively described in previous papers of our group (see \citealt{martocchia2018a}, \citealt{saracino2020} as examples), so we refer the interested reader to these papers for more details.

\begin{figure}
    \centering
	\includegraphics[width=0.46\textwidth]{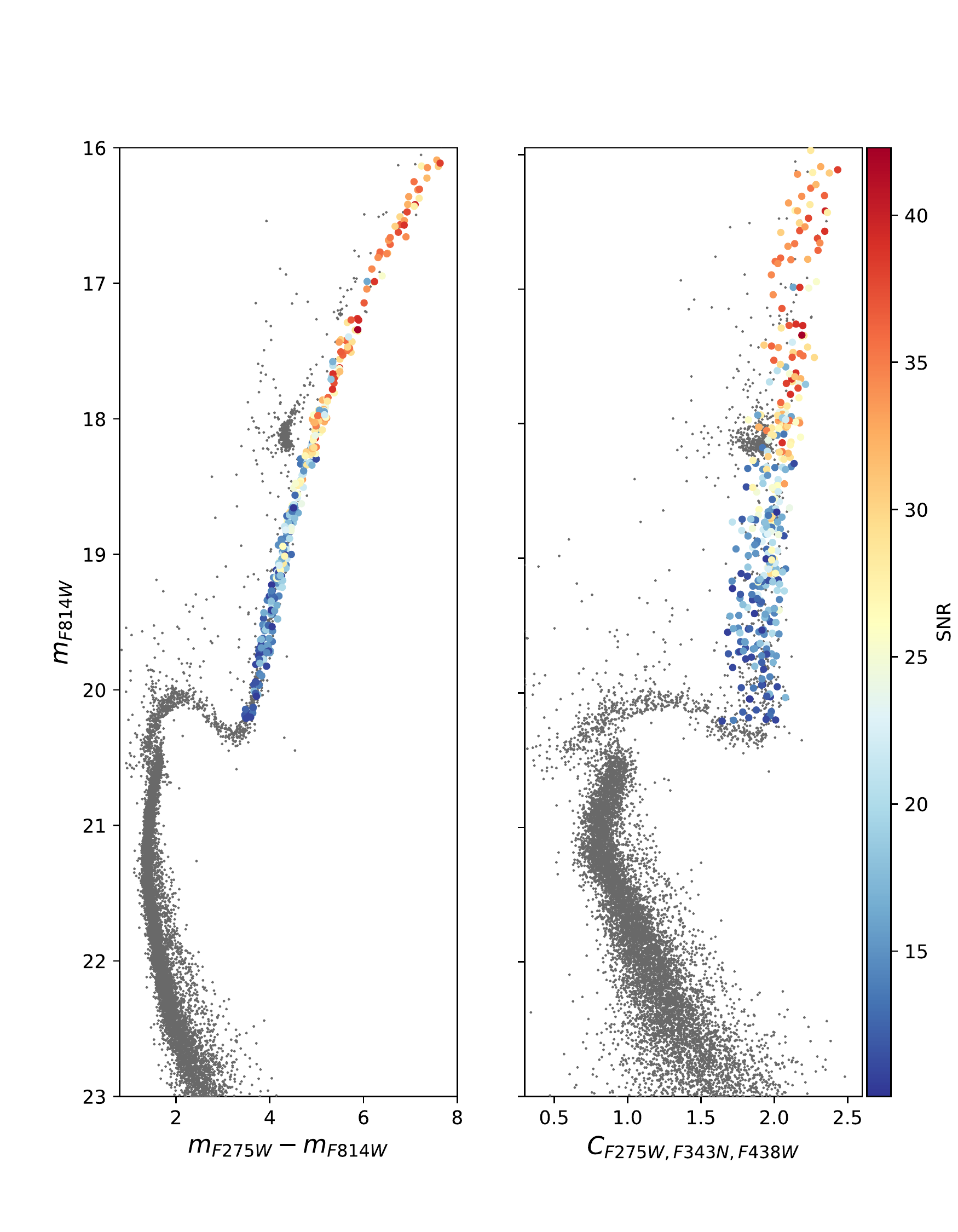}
    \caption{{\it Left panel:} ($m_{F814W}$, $m_{F275W}-m_{F814W}$) CMD for all the stars of NGC 1978 used in this work. {\it Right panel:} ($m_{F814W}$, $C_{F275W,F343N,F435/8W}$) CMD for the same stars. MUSE spectroscopic targets are overplotted as larger points, color-coded according to their quality in terms of signal-to-noise (${\mathrm S/N}$) ratio (see Section \ref{sec:cmap} for details).}
    \label{fig:cmd}
\end{figure}

By adopting the statistical approach explained in \citet[see also \citealt{cabrera-ziri2020}]{saracino2020}, we cleaned the color-magnitude diagram (CMD) of NGC~1978 from field star interlopers. Since no parallel fields are available in the HST archive for the cluster, we defined as cluster region, the area within 40'' of the cluster center ((RA, Dec) = (82.1860\textdegree, -66.2363\textdegree)), while as a control field region, we chose the area located at a distance greater than 75'' from the center. The procedure revealed the field star contribution to be negligible. The effect of the differential reddening has been also estimated, following the technique explained in \citet{milone2012}. The resulting $\delta E(B-V)$ are very low (on average around zero and with a maximum variation comparable to the photometric errors).
The ($m_{F814W}$, $m_{F275W}-m_{F814W}$) CMD of the innermost 40'' of NGC~1978 is presented in the left panel of Fig.~\ref{fig:cmd}, where the hook after the main sequence turn-off, at the sub giant branch (SGB) level, typical of a young star cluster, is clearly visible. In the right panel the pseudo-colour $m_{F814W}$ vs. $C_{F275W,F343N,F438W}$ diagram is shown. This filter combination is commonly used in MP studies since it is very powerful in separating stars with different N-abundances \citep[e.g.,][]{milone2017,lardo2018}. This filter combination will be used later on in Sec.~\ref{sec:cmap} to create the chromosome map of the cluster.
\vspace{-0.4cm}
\subsection{MUSE spectroscopy}

MUSE observations of NGC~1978 were obtained during two nights, 2019-11-29 and 2019-12-26, in the course of program 0104.D-0257 (PI: Kamann). Each night four $660\,$s exposures were observed. In between individual exposures, derotator offsets of $90^{\circ}$ were applied. The seeing during the observations (as measured on the reduced data) was typically $0.8\arcsec$. The location of the MUSE pointing is shown in Fig.~\ref{fig:map}.

The data were reduced with the standard MUSE pipeline \citep{weilbacher2020}. It performs all the basic reduction steps (bias removal, spectrum tracing, flat fielding, and wavelength calibration) on a per-IFU basis. Afterwards, the data from the 24 IFUs are combined and sky subtraction is performed. In the final step, all exposures taken during a single night are combined into a final data cube.

We used \textsc{PampelMuse} \citep{kamann2013,pampelmuse} to extract individual spectra from the MUSE cubes. As reference catalogue for the extraction, we used the HST data presented in Sec.~\ref{sec:obs:hst} above. In total, we obtained spectra for $4\,425$ stars. In this paper, however, we focus on the $338$ red giant stars (RGB) that have sufficient photometry to place them in the chromosome map (cf. Fig.~\ref{fig:cmap}) and were extracted with a spectral S/N $>$ 10 from the MUSE data. Their position in the CMD of NGC~1978 is shown in Fig.~\ref{fig:cmd}.

We processed the extracted spectra with \textsc{Spexxy} \citep{husser2016}, which determines radial velocities, metallicities, and stellar parameters via full spectrum fits against synthetic \textsc{GLib} \citep{husser2013} templates. Radial velocities and metallicities of the 338 stars used in this work are listed in Table \ref{tab:n1978}, along with their photometric informations. As explained in \citet{husser2016}, measuring the surface gravity from MUSE spectra is challenging. Instead, we determined $\log g$ for the extracted spectra via comparison of the HST photometry and the isochrone introduced in Sec.~\ref{sec:obs:models} below.
%\vspace{-1.58cm}
\subsection{Synthetic models}\label{sec:obs:models}
We calculated synthetic spectra and photometry in the same manner as \citet{martocchia2017} which is based on the earlier work of \citet{2012A&A...546A..53L}.
We refer the interested reader to that paper for further details.
From a 2.24 Gyr, [Fe/H]$ = -0.50$ MIST isochrone \citep[version 1.2][]{2011ApJS..192....3P, 2016ApJS..222....8D, 2016ApJ...823..102C}, we selected 35 evolution points in $\log$ $T_{eff}$-$\log L$ space between the start of the SGB and the tip of the RGB\footnote{Evolutionary phases where the temperature changes rapidly are better sampled. This works out to an spacing in log $T_{eff}$ of $\sim$0.008 dex and $\sim$0.015 dex on the RGB and the SGB, respectively.}. 
For each of these points we calculated a model atmosphere using \textsc{ATLAS12} \citep{1970SAOSR.309.....K, 2005MSAIS...8...14K} before using \textsc{SYNTHE} \citep{1979SAOSR.387.....K, 1981SAOSR.391.....K} to synthesize a spectrum for each\footnote{Both a model atmosphere and a synthetic spectrum were calculated for each of the three chemical compositions at each of the isochrone points.}.
Using the formula in \citet{2014MNRAS.444..392C} and the system response curves and zeropoints provided by the WFC3 website\footnote{ http://www.stsci.edu/hst/instrumentation/wfc3/data-analysis/photometric-calibration/uvis-photometric-calibration; see also \citet{deustua2016}}, we calculated synthetic photometry in a range of WFC3 filters from each model spectrum.  Synthetic photometry refers here to photometric models having as input parameters those used to compute the synthetic spectra.
Similar models have been used in previous works \citep{martocchia2017, martocchia2018a}, showing good qualitative agreement between the model magnitudes and observations
These models have been computed for three chemical compositions assuming the \citet{2009ARA&A..47..481A} solar abundance scale. In particular: {\it i)} a model with enhanced N and depleted C and O but scaled solar Na ([N/Fe] $= +0.5$, [C/Fe] = [O/Fe] $= -0.1$, [Na/Fe] = $0$); {\it ii)} \& {\it iii)} models with the same enhanced N and depleted C and O but two different Na enhancements ([Na/Fe] $= +0.1$ and [Na/Fe] $= +0.2$). Note that all three chemical compositions include enhanced N abundances. This choice is motivated by the predicted impact of the first dredge-up on the surface abundances of RGB stars. For all of the models we kept the He abundance and all other abundances constant at their scaled solar values and assumed that the surface abundances do not vary with stellar evolution. This last point is likely a valid assumption for Na, but C, N an O are affected by the first dredge up in clusters of this age \citep{salaris2020}.
Our models assume local thermal equilibrium.
\citet{2018ApJ...854..139C} studied the effect of non-local thermal equilibrium (NLTE) on the strength of the NaD lines in the context of integrated light studies.
They found it to be relatively small compared to the effect of [Na/Fe].
Furthermore, \citet{2018ApJ...854..139C} noted that the difference in the strength of a spectra feature with a change in abundance is less sensitive to the effects of NLTE than the absolute strength of the line since the effects of NLTE largely cancel. 
As a final note, we also mention that NLTE corrections depend on the stellar parameters and, as we show in Appendix \ref{app:tests}, the cumulative luminosity distributions of N-normal and N-enhanced stars are nearly identical. Thus, NLTE effects cancel out since we consider relative and not absolute differences.
%\vspace{-0.3cm}
\section{Results}\label{sec:res}
\subsection{Chromosome map of NGC 1978}\label{sec:cmap}
\begin{figure}
    \centering
	\includegraphics[width=0.47\textwidth]{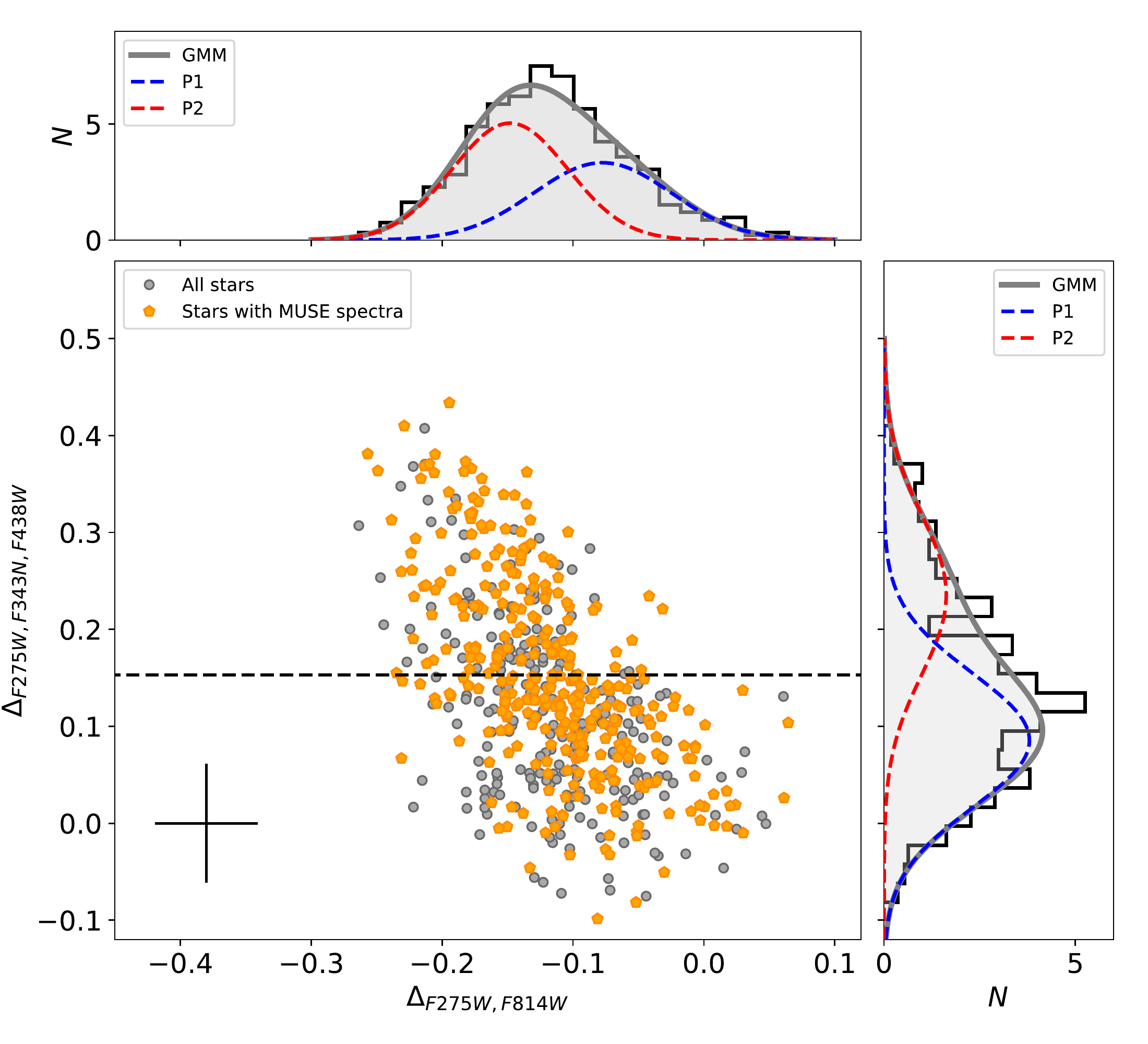}
    \caption{Chromosome map ($\Delta_{F275W,F814W}$, $\Delta_{F275W,F343N,F438W}$) of NGC~1978 (light-grey points). Orange pentagons are all the stars with extracted MUSE spectra, having $S/N > 10$. The black dashed line represents the separation adopted between the two populations of NGC 1978. {\it Top and right panels:} The histograms of the verticalised colour distribution $\Delta_{F275W,F814W}$ and of the verticalised pseudo-colour distribution $\Delta_{F275W,F343N,F438W}$ of NGC 1978, respectively. The solid grey line in both panels represents the result of the GMM fit, while blue and red dashed lines identify the Gaussian components related to P1 and P2, respectively (see the text for details).}
    \label{fig:cmap}
\end{figure}

NGC 1978, $\sim$2 Gyr-old, is the youngest cluster to date showing multiple stellar populations in the form of N spread. This result, discovered for the first time by \citet{martocchia2018b} via the pseudocolor $C_{F343N,F438W,F814W}$ diagram, has been further confirmed over the years using the chromosome map and different filter combinations (see \citealt{milone2020}). In the latter, the authors also photometrically inferred the abundances of C, N, O and He for the two populations of stars in NGC 1978, finding $\Delta$ [C/Fe] = -0.05$\pm$0.05, $\Delta$ [N/Fe] = 0.07$\pm$0.03, $\Delta$ [O/Fe] = 0.00$\pm$0.03 and $\Delta$ $Y_{max}$ = 0.002$\pm$0.003 (not accounting for the effect of the first dredge-up).
In this study, as already done for other LMC/SMC clusters in our survey \citep{saracino2019,saracino2020}, we exploited the power of the HST narrow band F343N, which is extremely sensitive to N differences between stars within the cluster, in combination with the near-UV F275W, which has been extensively adopted to detect MPs in Milky Way GCs (see \citealt{milone2017}). Using both diagrams presented in Fig.~\ref{fig:cmd}, we created the chromosome map of the RGB stars of NGC 1978, by adopting the same method (apart for negligible details) as the one defined by \citet{milone2015,milone2017} (see also \citealt{saracino2020}). Briefly, we computed the red and blue RGB fiducial lines in both color combinations, as the 5th and 95th percentile of the distribution, respectively. We then verticalised the distribution of RGB stars and normalized them for the intrinsic RGB width at 2 mag brighter than the turn-off, thus creating $\Delta_{F275W,F814W}$ and $\Delta_{F275W,F343N,F438W}$. These two quantities are plotted one against the other to produce the chromosome map of NGC 1978, which is presented in the main panel of Fig.~\ref{fig:cmap}, as grey dots. The distributions along both axes are shown as black histograms in the side panels of the figure. % \ref{fig:cmap}. 
The size of the photometric uncertainties (shown as a cross symbol in the bottom-left side of the figure), as well as the shape of the distribution itself confirm the presence of two populations of stars in the cluster. The first population, called P1 hereafter, is characterized by stars with solar-scaled N abundance, and is located in the bottom-right part of the plot, while the second population, P2 hereafter, is populated by N enhanced stars and occupies the top-left region. In agreement with what has been found for other intermediate-young age LMC/SMC clusters, the chromosome map of NGC 1978 does not show two well separated clumps, thus the distinction between P1 and P2 stars is not as clear as in some of the ancient Milky Way GCs \citep[e.g.,][]{milone2017}. It is worth to mention here that such a behaviour is not surprising for young clusters, being a direct consequence (at least partially) of the first dredge up \citep{salaris2020}. 

In order to explore the presence of N-normal and N-enhanced stars in NGC~1978 and to estimate the number ratios between these sub-populations, we applied a 1-dimensional Gaussian Mixture Model (GMM, \citealt{muratov2010}) analysis on the unbinned sample of $\Delta_{F275W,F814W}$ and $\Delta_{F275W,F343N,F438W}$.% The latter is especially interesting since it is a proxy of MPs.
The result is shown in the top and side panels of Fig.~\ref{fig:cmap} and it demonstrates the presence of two components, where the first is much more populated than the second. Blue and red lines in the panels refer to P1 and P2, respectively, while the grey solid line represents the combination of the two Gaussians as a result of the GMM fit. From the areas under the Gaussian functions we computed the number ratios between the sub-populations, finding $N_{P1} /N_{TOT}$ $\approx$ 0.66 $\pm$ 0.03 and $N_{P2} /N_{TOT}$ $\approx$ 0.34 $\pm$ 0.03, where $N_{TOT}$ is the total number of stars in the sample and 0.03 is the typical uncertainty of the number ratios estimated via bootstrapping. Photometric errors, residual differential reddening, inclination angle of the populations with respect to the y-axis, etc. can indeed have an impact on such percentages. We performed the same analysis by exploiting the 2-dimensional GMM, which mostly gives the same separation between P1 and P2, except for a handful of stars at the limit between the two populations (see Appendix \ref{app:tests} for further details).

%\vspace{-0.5cm}
\subsection{Sodium variations between P1 and P2 stars}
\label{sec:results:na}

To constrain the element abundances between P1 and P2 stars, we created combined MUSE spectra of both populations. This was done following \citet{latour2019}, where a similar analysis has been performed on the Galactic GC NGC~2808 (see their section 3.2).
First, we identified the P1 and P2 stars in the MUSE spectral sample. As the two sub-populations are not clearly separated, we chose a simple method of splitting the population based on the $\Delta_{F275W,F343N,F438W}$ pseudo-colour, which is shown as a dashed line in Fig.~\ref{fig:cmap}. In order to remove possible field stars that survived the photometric cleaning, we removed all stars from our samples for which the spectral analysis resulted in a metallicity outside the interval $-1.0 < [{\rm M/H}] < -0.2$ or a radial velocity outside the interval $285\,{\rm km\,s^{-1}} < v_{\rm r} < 305\,{\rm km\,s^{-1}}$. In addition, spectra extracted with ${\rm S/N}<20$ were omitted, resulting in final sample sizes of 244 spectra of P1 stars and 145 spectra of P2 stars. Each of the remaining spectra was divided by the telluric component obtained during its spectral analysis. Further, each spectrum was divided by the polynomial that \textsc{Spexxy} determined in order to account for possible continuum mismatches between the observed spectrum and the synthetic templates. Finally, the spectra of both populations were averaged. Note that, as in \citet{latour2019}, we did not weigh the individual spectra by their S/N during the combination. Furthermore, in light of the small expected velocity dispersion of NGC~1978 ($\lesssim 5\,{\rm km\,s^{-1}}$), we did not correct the individual spectra for their measured radial velocities.

Table \ref{tab:n1978} contains photometric information for all the MUSE targets included in this work, along with the S/N of the individual spectra, and the adopted separation between P1 and P2 stars. We stress here that the results shown later in the Section do not depend on the adopted separation between P1 and P2. Indeed, comparing only the most extreme P1 and P2 stars (i.e. those located at the edges of the $\Delta_{F275W,F343N,F438W}$ distribution) results in a very similar answer. This and other tests are presented in detail in Appendix \ref{app:tests} at the end of the paper.

\begin{figure*}
    \centering
    \includegraphics[width=0.98\textwidth]{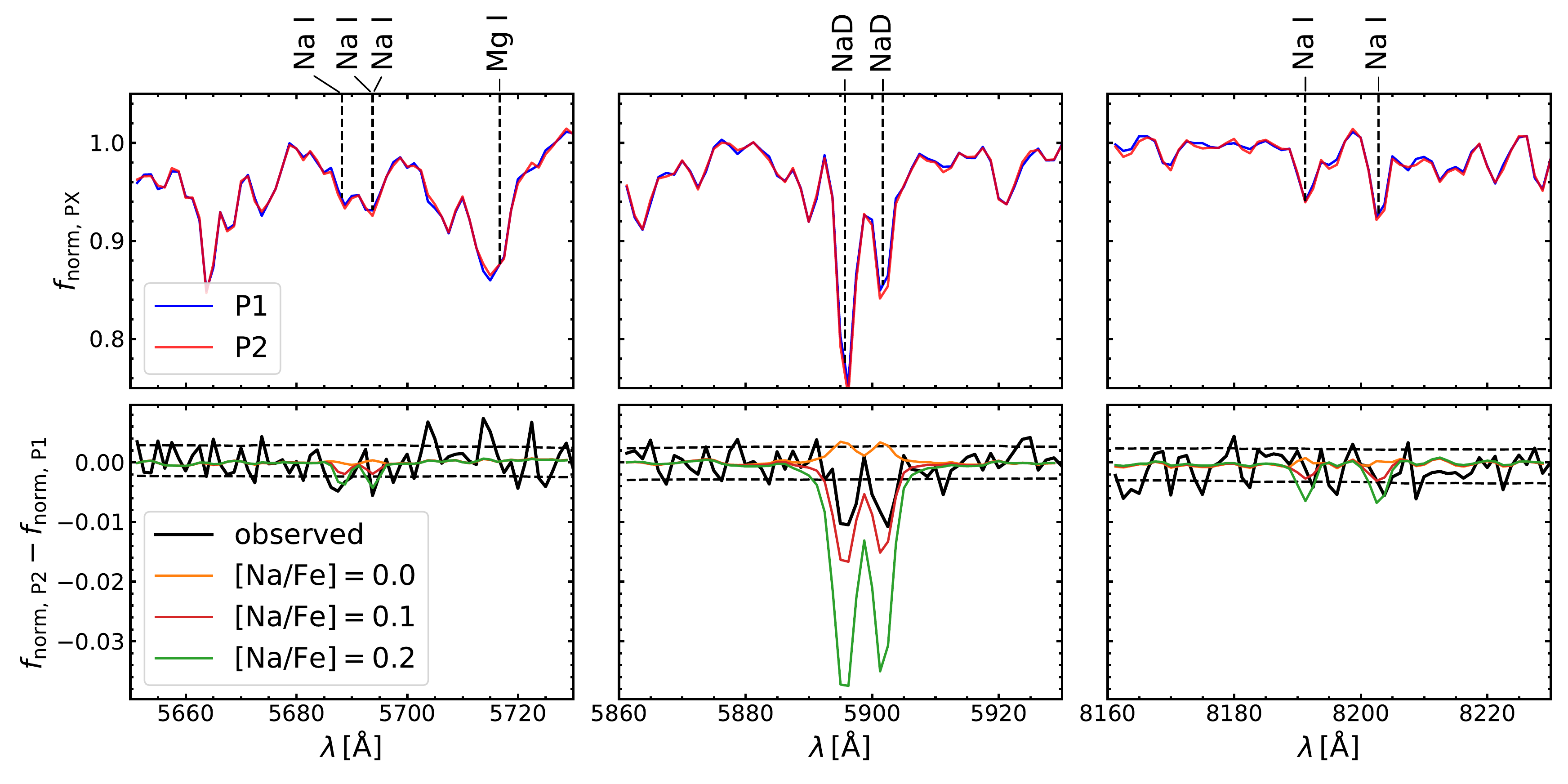}
    \caption{\textit{Top panel:} Normalized combined spectra created for populations P1 and P2 around different sodium lines covered in the MUSE spectral range. \textit{Bottom panel:} The flux difference between the normalized spectra of P1 and P2 is shown as a solid black line. Dashed black lines indicate the $1\sigma$ scatter of the flux difference. The expected differences for increasing sodium enrichment in P2 are shown as solid coloured lines.}
    \label{fig:spec:na}
\end{figure*}

We normalized the combined spectra of P1 and P2 by dividing them through their best-fit continua. The continua were determined by splitting each spectrum into 25 equally-sized wavelength bins, calculating the median of the 15\% highest pixel values in each bin, and performing a second-order spline interpolation between these values. In Fig.~\ref{fig:spec:na}, we compare the resulting spectra of P1 and P2, focusing on spectral regions containing Na lines. The top row of Fig.~\ref{fig:spec:na} directly compares the normalized spectra of the two populations whereas in the bottom row, we show the residuals after subtracting the spectrum of P1 from the one of P2. The dashed lines in the bottom panels show the $1\sigma$ scatter of the residuals, obtained by calculating the running standard deviation with a window size of 200 spectral pixels.

Only for the strong NaD doublet ($5\,990\,{\text \AA}$, $5\,996\,{\text \AA}$), we can see a significant difference in Fig.~\ref{fig:spec:na}, with P2 showing stronger lines than P1. We converted the observed difference into an equivalent width by fitting a double Gaussian to the data shown in the bottom centre panel of Fig.~\ref{fig:spec:na}. The centroids of the Gaussians were set to the known wavelengths of the two NaD lines and a common FWHM of $2.5\,{\text \AA}$ (corresponding to the spectral resolution of MUSE) was adopted. Then, their areas were determined in a linear least squares fit. Furthermore, we estimated uncertainties for our measurements by adding random noise to the observed difference, according to the estimated scatter in the residuals (i.e. the dashed lines visible in Fig.~\ref{fig:spec:na}). This resulted in a difference of $-66\pm9\,{\rm m}{\text \AA}$. An additional approach has been also used to determine the uncertainties of our measurements. This method is described in detail in Appendix \ref{app:sigma}, confirming our previous findings.

We further searched for differences in other spectral regions, paying particular attention to elements that are known to vary in Galactic GCs (like Mg, Al, or O). For an overview of the relevant lines covered by MUSE, see \citet{latour2019}. However, no significant differences were detected. In particular, the strong Mg triplet lines ($5\,167\,{\text \AA}$, $5\,173\,{\text \AA}$, $5\,184\,{\text \AA}$) are indistinguishable between the combined spectra of P1 and P2 (see Appendix~\ref{app:elements} and Fig.~\ref{fig:mg_comp}).

Finally, we also verified that binary stars do not impact our analysis. Making use of the two epochs of MUSE data, we investigated the radial velocity variations of the P1 and P2 stars using the method developed by \citet{giesers2019}. Both populations appear to have low binary fractions, $\lesssim5\%$. While the binary fraction of P1 appears to be slightly enhanced compared to P2, further analyses based on additional data will be required to verify this.

\begin{table*}
\caption{Photometric and spectroscopic information of the RGB MUSE targets used for the analysis of NGC~1978. The full Table will be available in the online version of the paper.}
\begin{tabular}{c c c c c c c c c c}
\hline
SourceID & RA & Dec &  $\Delta_{F275W,F814W}$ & $\Delta_{F275W,F343N,F438W}$ & $[Fe/H]$ & $R_{vel}$ & SNR & N$_{exp}$ & Pop. \\  
\hline
1 & 82.1776505 & -66.2284622 & -0.067 & 0.018 & -0.64$\pm$0.02 & 292.16$\pm$1.28 & 29.08 & 2 & P1 \\    
2 & 82.1704636 & -66.2304230 & -0.172 & 0.139 & -0.64$\pm$0.09 & 293.47$\pm$3.57 & 11.28 & 2 & P1 \\   
3 & 82.1743088 & -66.2291718 & -0.046 & 0.149 & -0.66$\pm$0.03 & 294.88$\pm$1.90 & 17.89 & 2 & P1 \\    
$\dots$ & $\dots$ & $\dots$ & $\dots$ & $\dots$ & $\dots$ & $\dots$ & $\dots$ & $\dots$ & $\dots$ \\
195 & 82.1713028 & -66.2295151 & -0.192 & 0.324 & -0.67$\pm$0.02 & 288.62$\pm$1.06 & 32.49 & 2 & P2 \\
196 & 82.1729279 & -66.2301025 & -0.120 & 0.185 & -0.59$\pm$0.06 & 293.04$\pm$2.69 & 14.18 & 2 & P2 \\
197 & 82.178093  & -66.2306595 & -0.154 & 0.177 & -0.65$\pm$0.02 & 297.27$\pm$0.94 & 33.04 & 2 & P2 \\
$\dots$ & $\dots$ & $\dots$ & $\dots$ & $\dots$ & $\dots$ & $\dots$ & $\dots$ & $\dots$ & $\dots$ \\
\hline
\end{tabular}
\label{tab:n1978}
\end{table*}

%\vspace{-0.5cm}
\subsection{Comparison with synthetic spectra}

In order to examine the differences we found for the NaD lines as a variation in Na between P1 and P2, we made use of the theoretical models introduced in Sec.~\ref{sec:obs:models}. We proceeded by selecting for each star with a valid MUSE spectrum the template spectrum of the model star closest in terms of F814W magnitudes. Stars in P1 were only matched to the models with chemistry \textit{(i)}, whereas stars in P2 were matched to the models with chemical compositions \textit{(i)} to \textit{(iii)}. We combined the mock spectra in the same way as we did for the observed spectra and finally calculated the expected differences between P1 and P2 for all three P2 chemical compositions listed in Sec.~\ref{sec:obs:models}. These expected differences are included in the bottom panels of Fig.~\ref{fig:spec:na}.

We can see in the bottom centre panel of Fig.~\ref{fig:spec:na} that in the case that both populations share the same Na abundance, one would expect slightly stronger NaD lines in the combined spectrum of P1. This is likely caused by slightly different effective temperature distributions in the MUSE samples of P1 and P2. With increasing Na abundance in P2, however, negative NaD residuals in the difference spectrum are predicted, with the observed difference being similar to the model prediction for a Na enrichment of $+0.1\,{\rm dex}$. We converted the predicted differences in the NaD lines into equivalent widths in the same way as we did for the observed spectra (cf. Sec.~\ref{sec:results:na}). Then, we linearly interpolated the model results to our measured difference of $-66\pm9\,{\rm m}{\text \AA}$. This method yields a sodium enrichment of P2 of $[{\rm Na/Fe}]=0.07\pm0.01$ dex. The uncertainty of our Na difference measurement was determined by running the same analysis for the random samples mentioned in Sect.~\ref{sec:results:na}.
Finally, the bottom left and bottom right panels of Fig.~\ref{fig:spec:na} show that for the level of Na enhancement inferred from the NaD lines, any expected differences in the weaker Na lines covered by MUSE are below our detection limit.

%\vspace{-0.5cm}
\section{Discussion and Conclusions}\label{sec:concl}
In this work we combined HST/WFC3 photometry with VLT/MUSE spectroscopy of a sample of 338 RGB stars in NGC~1978, the youngest ($\sim2$~Gyr) cluster to date with reported MPs \citep{martocchia2018b}, to look for light element (e.g., Na, Al, Mg and O) abundance variations among its populations. 

Indeed, while the presence of N spreads had been widely observed in clusters younger than 10~Gyr lying in external galaxies (e.g., Magellanic Clouds), no evidence for other element variations (e.g., Na) has been detected so far in these systems. This raised the question of whether the abundance variations detected in such young clusters could have a different origin than those of their older counterparts in the Galaxy and in the MCs. 

This paper is the first of a series aimed at shedding new light on this crucial aspect, focusing on elements (e.g. Na) which are almost insensitive to mixing effects (i.e., the first dredge-up, \citealt{salaris2020}).

Being inspired by the recent work of \citet{latour2019}, we made use of the ``chromosome map’’ as a powerful tool to distinguish N-normal from N-enhanced RGB stars in the LMC cluster NGC~1978. Among the MUSE targets, we identified those belonging to each component and combined their spectra to get one spectrum with high S/N for each population. The spectra of P1 and P2 clearly show a different strength in the NaD lines, with P2 being enriched by $[{\rm Na/Fe}]=0.07\pm0.01$ dex with respect to P1. Such a variation is unfortunately under our detection limit for the weaker Na lines, as well as for other spectral lines (e.g. Mg, Al, O). Such small variations may explain why earlier work by \citet{mucciarelli2008} did not find evidence for significant Na spreads within the cluster. 

When compared to the Na abundance variations obtained by \citet{latour2019} for the ancient Galactic GC NGC~2808 (see their Table 3), the $[{\rm Na/Fe}]$ enrichment derived here appears significantly smaller. However, there are manifold reasons why these two clusters are difficult to compare directly: \textit{1)} NGC 2808 and NGC 1978 have different masses, with the LMC cluster being lighter than the Galactic GC. Since the [Na/Fe] variations scale positively with the absolute luminosity, i.e. the cluster mass, in ancient MW GCs \citep{carretta2014} and the same has been observed also for other elements (e.g. N), such difference in their Na spreads is readily expected. \textit{2)} NGC 2808 and NGC 1978 have very different metallicities ([Fe/H]=-1.15 and [Fe/H]=-0.5, respectively). \textit{3)} NGC~2808 is an extreme case, hosting at least five distinct populations \citep[e.g.,][]{milone2015}. If only the P1 and P2 populations of NGC~2808 are included in the comparison ($\Delta$[Na/Fe]$=0.14\pm0.06$) then the difference is greatly reduced but the results are nevertheless different. Similar findings come from the comparison of the $[{\rm Na/Fe}]$ difference derived in this work for NGC 1978 to that listed in Table 2 by \citet{marino2019} for some Galactic GCs having, at least, similar masses: the observed sodium difference is systematically higher in the ancient Galactic GCs compared to what we find here for NGC 1978. 
If there is a positive correlation between the sodium abundance variation and the cluster age (not yet explored), then it becomes reasonable that MPs in NGC1978 have a much lower sodium difference compared to older GCs. See \citet{martocchia2018b} and \citet{lagioia2019} for different conclusions on the presence of an age trend for nitrogen.

Chromosome maps are not used here to support the comparison between old and young clusters as we know that the mixing effect caused by the first dredge-up makes the interpretation of such a diagnostic in terms of N abundance variations more complicated \citep{salaris2020}.

While the correlation between age and abundance spreads will be further explored in the coming papers of this series, the Na-abundance variation detected here between P1 and P2 comes as a new and independent evidence that the MP phenomenon shows the same features, regardless of cluster age and host galaxy.

We have shown that the MPs in this young cluster are effectively the same as in the ancient GCs.  Hence, young clusters can be used to place stringent constraints on the origin of MPs.  First, as they are not restricted to only the ancient GCs, they cannot be linked to the special conditions of the early Universe \citep[e.g.,][]{kruijssen15}. Additionally, mechanisms that only work at low metallicity also appear to be in conflict with these (and other) observations. Due to their youth, young clusters like NGC~1978 and NGC~2121 can be used to search for evidence of multiple epochs of star-formation (i.e. multiple generations) within them, and if they exist, what is the age difference between the generations? \citet{martocchia2018b} and \citet{saracino2020} have shown that the two populations within NGC~1978 and NGC~2121, respectively, are coeval, within the uncertainties, with an upper limit of $\sim15-20$~Myr.  This is in conflict with theories of multiple generations which invoke AGB stars as the source of the processed materials.  These models require $\sim30-100$~Myr to begin operating.

\section*{Data Availability}
The data underlying this article are available in the article and in its online supplementary material.

%\vspace{-0.7cm}
\section*{Acknowledgements}
We thank the referee for his/her detailed review, which helped to strengthen the results.
We gratefully acknowledge Florian Niederhofer, Vera Kozhurina-Platais and S\o ren Larsen for helpful comments and discussion. SS, SK, NB and SM gratefully acknowledge financial support from the European Research Council (ERC-CoG-646928, Multi-Pop). NB also acknowledges support from the Royal Society (University Research Fellowship). This work is part of HST GO-14069 and GO-15630 programs and support for this work was provided by NASA through Hubble Fellowship grant HST-HF2-51387.001-A awarded by The Space Telescope Science Institute, which is operated by the Association of Universities for Research in Astronomy, Inc., for NASA, under contract NAS5-26555.
Based on observations collected at the European Southern Observatory under ESO programme 0104.D-0257.
\vspace{-0.5cm}
%%%%%%%%%%%%%%%%%%%%%%%%%%%%%%%%%%%%%%%%%%%%%%%%%%



\bibliographystyle{mnras}
\bibliography{lind1} % if your bibtex file is called example.bib

%\clearpage
\appendix
\section{Additional tests}
\label{app:tests}
In this appendix we collect all the additional tests we have performed to support the main scientific result of this work: the presence of a sodium abundance difference between P1 and P2 stars in the LMC cluster NGC 1978. This choice has been made to ensure a better readability of the whole text.
\subsection{Different definitions for P1 and P2 stars}
\begin{figure}
    \centering
	\includegraphics[width=0.46\textwidth]{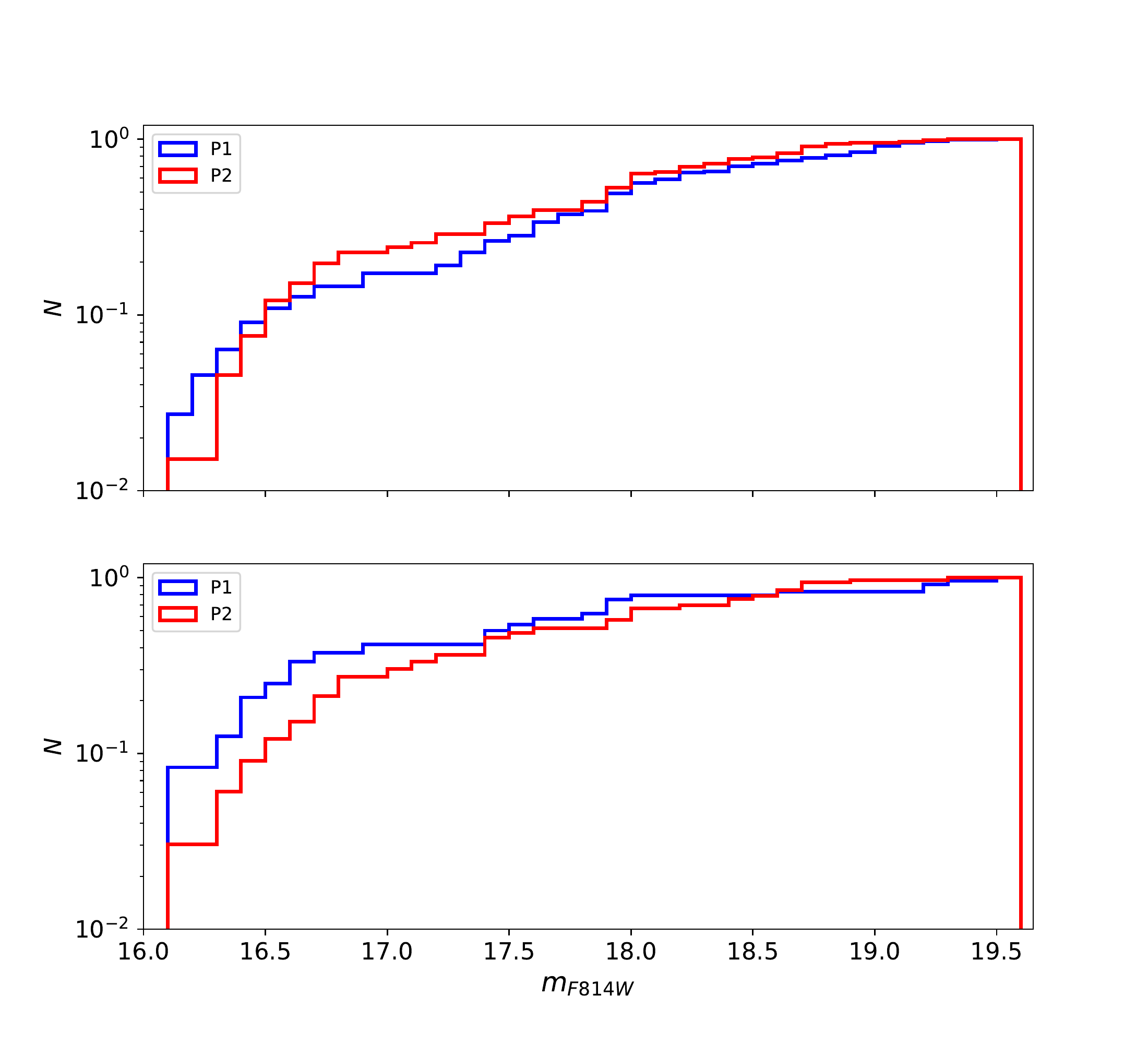}
    \caption{Normalized cumulative distribution of P1 and P2 stars, as a function of $m_{F814W}$, is presented as blue and red histograms, respectively. Top panel for configuration 1. (P1 and P2 stars classified according to Figure \ref{fig:cmap}). Bottom panel for configuration 2. (P1 and P2 stars classified according to Figure \ref{fig:cmap2}). Although the number of stars is slightly different, the cumulative functions are roughly the same in both populations/configurations.}
    \label{fig:LF}
\end{figure}

For our spectroscopic analysis we have classified RGB stars belonging to P1 or P2 in NGC 1978 according to the dashed line shown in Figure \ref{fig:cmap}. This is not a random assumption since this value along the y-axis of the chromosome map corresponds to the locus where the (mean$_{P1}$-$\sigma_{P1}$) of the blue gaussian distribution intercepts the (mean$_{P2}$+$\sigma_{P2}$) of the red one. The parameters (mean and $\sigma$) of each population are the result of the GMM fit.
To corroborate such a choice, we computed the luminosity function of P1 and P2 stars thus selected as a function of the $m_{F814W}$ magnitude. As we can see in the top panel of Figure \ref{fig:LF}, where the cumulative distribution of P1 (in blue) and P2 (in blue) are plotted one on top of the other, they show roughly the same trend, regardless of the size of the two samples. This strongly limits the possibility that the sodium difference we see by comparing P1 and P2 stars could be driven by other effects like the effective temperature of the stars. 

\begin{figure}
    \centering
	\includegraphics[width=0.46\textwidth]{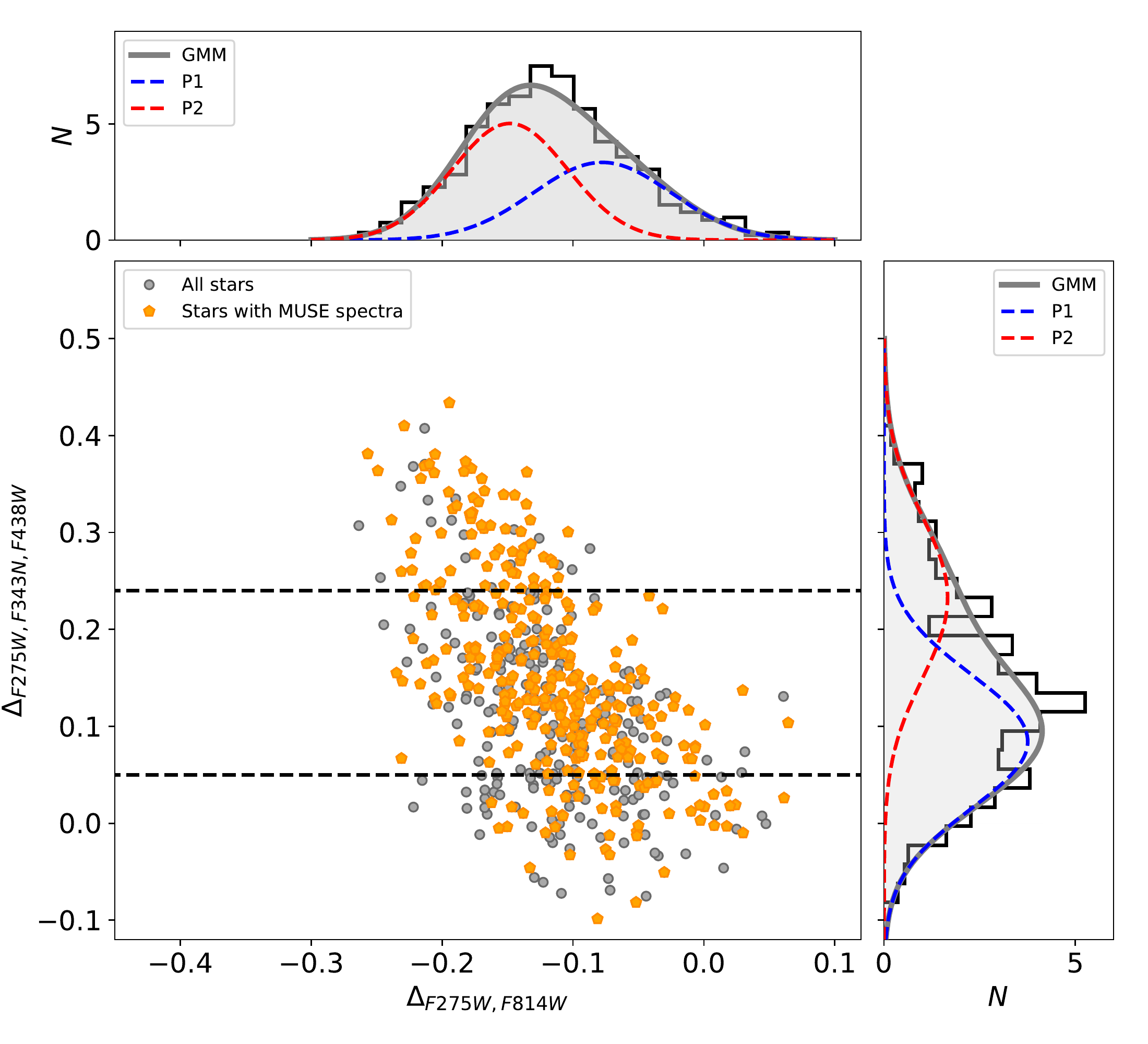}
    \caption{The chromosome map of NGC 1978 is shown as in Figure \ref{fig:cmap}, but the dashed lines shown here in the main panel represent the limits adopted for the new classification of P1 and P2 stars.}
    \label{fig:cmap2}
\end{figure}

The strongest test in favor of the detection of an internal [Na/Fe] spread within NGC 1978 comes from another classification of P1 and P2 stars. Indeed, starting from the chromosome map, we have considered as P1 and P2 stars only those RGB stars located at the two tails of the $\Delta_{F275W,F343N,F438W}$ distribution. In particular, P1 stars fall at $\Delta_{F275W,F343N,F438W} < 0.05$ and P2 stars at $\Delta_{F275W,F343N,F438W} > 0.24$ (see Figure \ref{fig:cmap2} for a visual inspection). This assumption has been made to exclude from the sample all those stars located in the overlapping region between the two sub-populations. In this way, the result of the comparison between the two samples is unlikely to be contaminated by stars wrongly associated with P1 instead of P2 and vice versa. 

\begin{figure}
    \centering
	\includegraphics[width=0.47\textwidth]{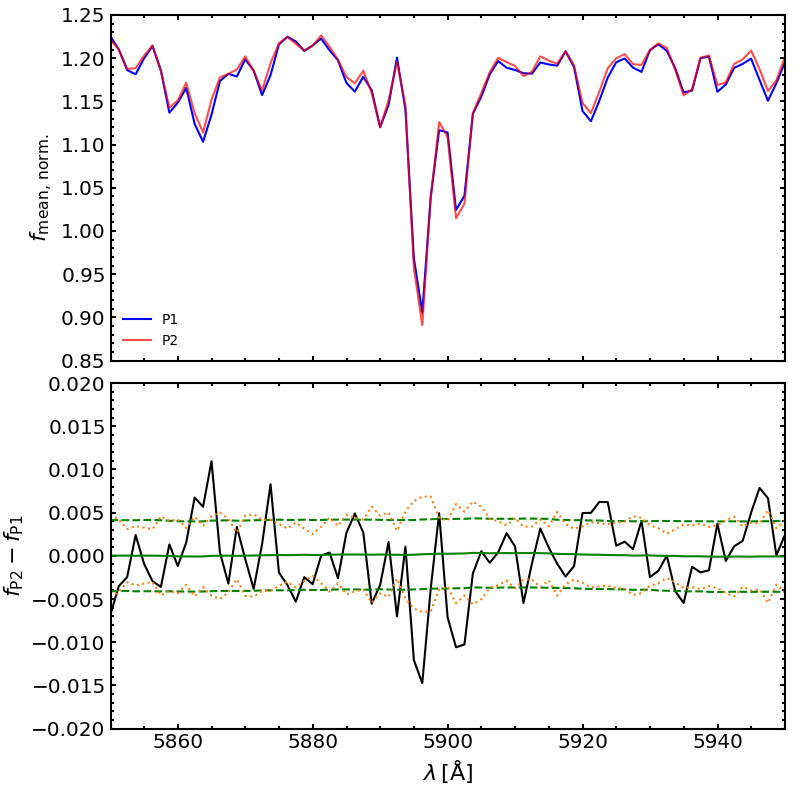}
    \caption{\textit{Top panel:} Normalized combined spectra of P1 and P2 around the NaD lines at $\sim$5900${\text \AA}$, for the classification shown in Figure \ref{fig:cmap2}. \textit{Bottom panel:} The flux difference between the normalized spectra of P1 and P2 is shown as a solid black line. Dashed black lines indicate the $1\sigma$ scatter of the flux difference. Orange dotted lines instead show the $1\sigma$ scatter of the flux difference computed via bootstrapping.}
    \label{fig:na_comp2}
\end{figure}

Although the size of the samples is significantly reduced, leading to a decrease of the S/N of the combined spectra, the signal of a sodium difference between the first and the second population is still clearly detectable (see Figure \ref{fig:na_comp2}). By adopting the approach described in Section \ref{sec:obs:models} for an appropriate comparison with synthetic models, we get a difference of $-79\pm23\,{\rm m}{\text \AA}$ which corresponds to $\Delta$ [Na/Fe] = 0.12$\pm$0.02. The significance of such a detection is a bit smaller but it is not surprising as both spectra are much noisier due to the lower S/N. It is worth noting, however, that the sodium difference we get from such a test is slightly higher compared to the previous analysis. This result is expected as the sodium abundance of each population is now only marginally affected by dilution due to possible interlopers. We have verified, also in this case, that the luminosity functions of the two smaller samples are still comparable to each other (see bottom panel of Figure \ref{fig:LF}). 

As a bottom line, we have also explored a sample selection based on a 2-dimensional GMM, in order to separate the two populations. As expected, using this method, relative to our preferred method (see Section \ref{sec:cmap}), the main differences were a handful of stars near the border between the two populations, which causes the Na-difference in terms of equivalent width to change from -66${\rm m}{\text \AA}$ to -61${\rm m}{\text \AA}$. We then conclude that the definition of the sample selection (at least those explored in this paper) does not significantly change our results.

\subsection{The impact of interstellar absorption on NaD lines}
\label{app:absorption}
It is well known that the NaD lines studied in this work can be affected by interstellar absorption. However, such absorption will only impact our results if its contribution to the combined spectra of P1 and P2 is different. This could be the case when (a) there is \emph{differential} interstellar absorption across our field of view and (b) the are statistical differences in the spatial distributions of the P1 and P2 stars. Already in Section \ref{sec:obs:hst}, we verified that differential reddening is negligible in the case of NGC 1978, at least in the field of view covered by our observations. However, it is still worth looking at the spatial distributions of the P1 and P2 stars with respect to the cluster centre, to see whether any residual differential interstellar absorption could have affected the spectra created for the two populations in a significantly different way.

The comparison between their radial distributions, performed via a 2-sided Kolmogorov-Smirnov test, reveals P2 stars to be more centrally concentrated than P1 stars, at a 3-sigma level. This result is in agreement with the findings of \citet{dalessandro2019}, who adopted $A^{+}$ (i.e. the area within the cumulative distribution functions of P1 and P2) as a parameter to classify concentration differences between the multiple populations within a cluster. To be sure that our results are not affected by the different concentrations, we statistically selected a subsample of P1 stars that has the same concentration as the P2 sample. This was done by selecting, for each star in P2, the closest P1 star in terms of distance to the cluster centre. We then performed the same differential spectral analysis between the combined spectra of the two samples as described in Sect.~\ref{sec:results:na}. We find a difference in the combined equivalent width of the NaD lines of $-54\pm16\,{\rm m}{\text \AA}$, fully consistent with the results obtained in Sect.~\ref{sec:results:na}. Hence we are confident that interstellar absorption does not affect the conclusions of our analysis.

\subsection{Significance of the Na spread within NGC 1978}
\label{app:sigma}
\begin{figure}
    \centering
	\includegraphics[width=0.47\textwidth]{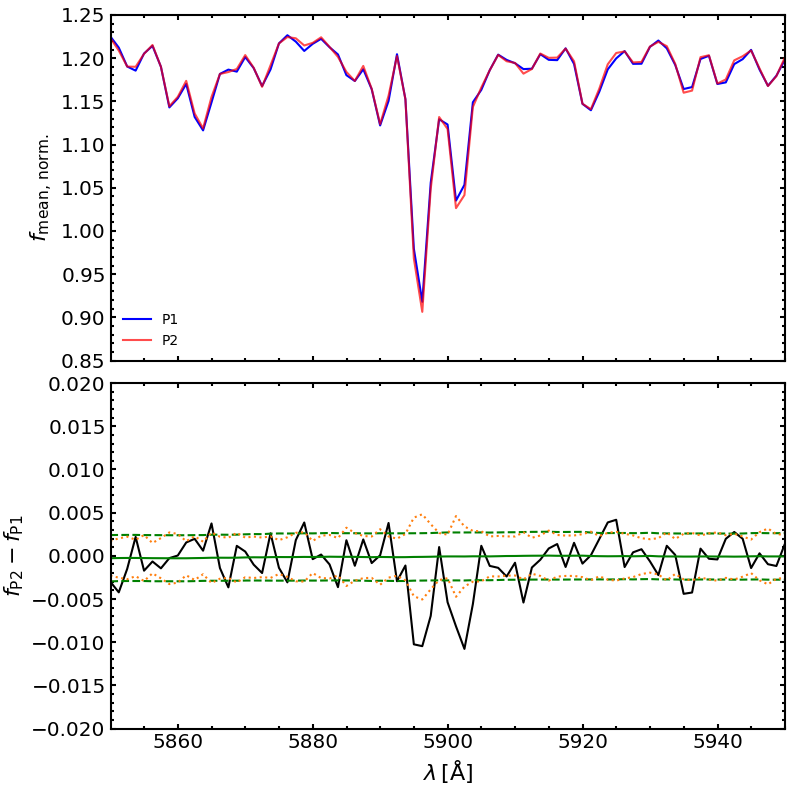}
    \caption{As in Figure \ref{fig:na_comp2} but for the classification of P1 and P2 defined in Figure \ref{fig:cmap}.}
    \label{fig:na_comp3}
\end{figure}

The sodium abundance spread observed in NGC 1978 is rather small, of the order of 0.1 dex. %, but we are confident it is pretty solid as it corresponds to a $\sim$3$\sigma$ detection. 
To give a confirmation that such a detection is significant, we have adopted a different method to infer the uncertainties of our measurements. As reliable uncertainties are not available for the individual MUSE spectra, we have applied a bootstrap technique to both selection methods (i.e. the method adopted in the main paper, Fig. \ref{fig:cmap}, as well as the extreme ends of the chromosome map in Fig. \ref{fig:cmap2}). Briefly, we created 50 bootstrap realizations of each combined spectrum (i.e. randomly selected $N$ spectra with repetitions from the parent sample of $N$ spectra) and measured their scatter as a function of wavelength. 
The new estimate of the noise is shown as orange dotted lines in the bottom panels of Figure~\ref{fig:na_comp2} and Figure~\ref{fig:na_comp3}, respectively, for the two selection methods used in the paper. As can be seen, the agreement with the previous estimate (black dashed lines) is remarkably good. However, we note that the noise increases by a factor of 1.5-2 at the position of the NaD lines. This increase in the noise directly reflects on the measurement uncertainty, which also increases by a factor of $\lesssim$2. By adopting this new approach, we can then conclude that, in both cases, the significance of the observed $\Delta$[Na/Fe] is somewhat smaller than before but still significant (to more than 3$\sigma$).

\subsection{Constraints on other elements}
\label{app:elements}

\begin{figure}
    \centering
    \includegraphics[width=0.47\textwidth]{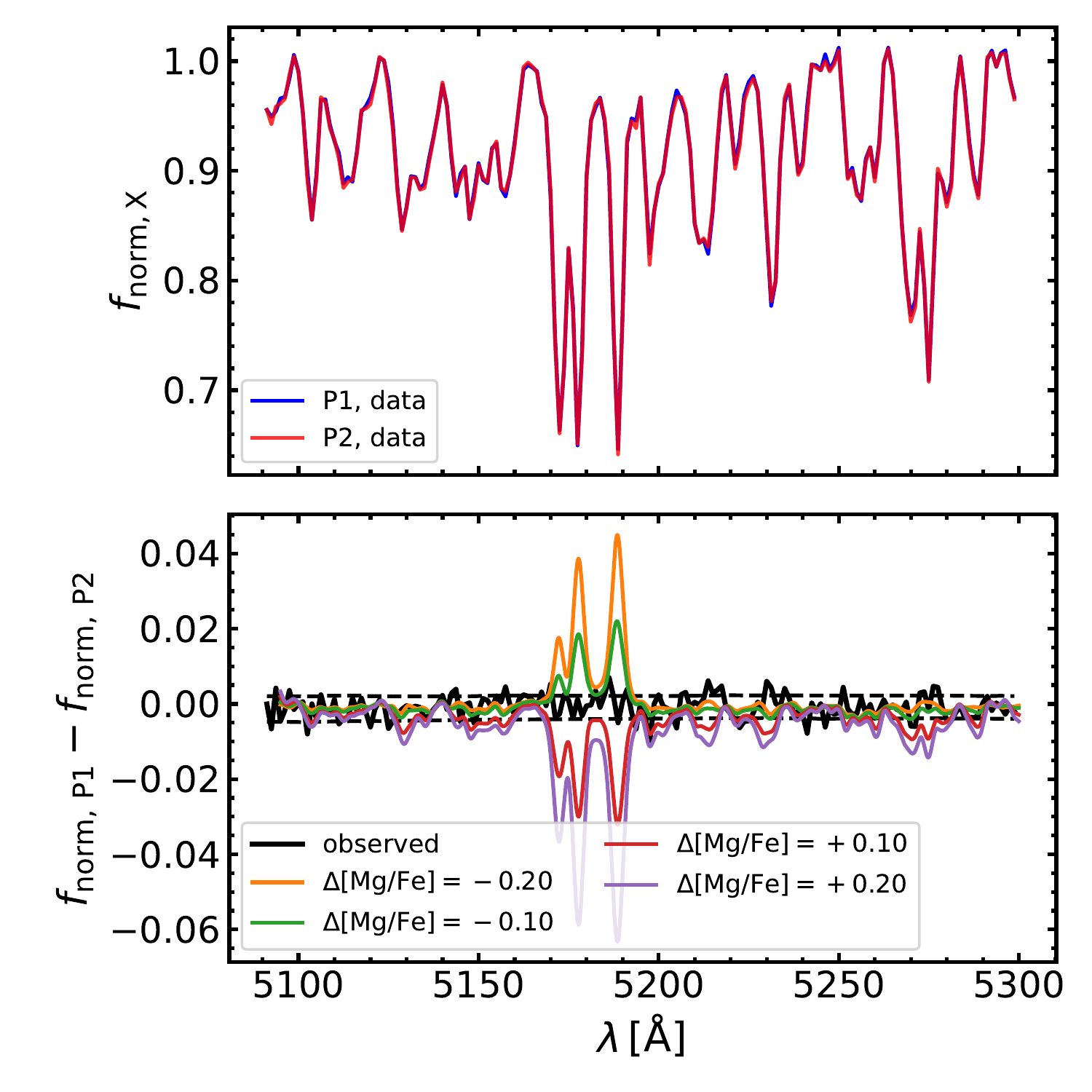}
    \caption{\textit{Top panel:} Normalized combined spectra of P1 and P2 around the Mg triplet. \textit{Bottom panel:} The flux difference between the normalized spectra of P1 and P2 is shown as a solid black line. Dashed black lines indicate the $1\sigma$ scatter of the flux difference. Coloured lines give the differences expected for various Mg abundance differences between P2 and P1.}
    \label{fig:mg_comp}
\end{figure}

As mentioned above, we did not find any variations between P1 and P2 when comparing the lines of other elements, such as Mg or Al. In Fig.~\ref{fig:mg_comp}, we illustrate this for the strong Mg triplet lines. We include in the bottom panel of Fig.~\ref{fig:mg_comp} the differences predicted for various levels of Mg enrichment or depletion of P2. These predictions were obtained by calculating the same synthetic spectra as in \citet{latour2019}, but for a metallicity of -0.5. It becomes clear that even differences of $\Delta[{\rm Mg/Fe}]$ of $\pm0.1\,{\rm dex}$ should be clearly visible in our analysis. Hence we conclude that no significant Mg spread exists in NGC~1978.

We applied the same approach to the Al and O lines that exist within the MUSE spectral range. As these lines are considerably weaker than the Mg triplet lines, our upper limits on the variation of Al or O between P1 and P2 are less stringent. We found that variations of $\Delta[{\rm Al/Fe}]\lesssim\pm0.1\,{\rm dex}$ and $\Delta[{\rm O/Fe}]\lesssim\pm0.2\,{\rm dex}$ would remain undetected given our current S/N limit.

The models of \citet{latour2019} further allowed us to perform how sensitive the derived Na abundance difference between P1 and P2 is on the adopted synthetic models. We find that the inferred Na difference is consistent between the two modelling approaches, giving us further confidence into the robustness of our result.

% Don't change these lines
\bsp	% typesetting comment
\label{lastpage}
\end{document}